\documentclass[12pt]{article}
\usepackage{graphicx}

\usepackage{authblk}

\usepackage{hyperref}

\usepackage{textcomp}

\date{\today}
\title{\bf Helium beam particle therapy facility}
\author[]{Mariusz Sapinski \footnote{mail: mariusz.sapinski@cern.ch}}
\affil[]{GSI, Darmstadt, Germany} 
\affil[]{SEEIIST, CERN, Geneva, Switzerland}

\begin{document}
\maketitle
\thispagestyle{empty}

\begin{abstract}

  Due to its precision and limited side effects, the particle therapy of cancer is gaining popularity.
  The number of patients treated with protons and light ions reached 150,000 worldwide.
  There are currently more than 80 facilities, which offer this treatment and
  several dozen of new ones are in construction.
  Mostly they are cyclotron-based facilities, which provide proton beams,
  and only several are synchrotron-based facilities that provide both: proton and carbon-ion beams.
  The advantage of carbon ions is their higher efficiency in destroying the cancer cells,
  more localized shape of the Bragg peak and smaller beam scattering in the body.
  Several of the large centers, which can provide carbon, have experimented with other ions including helium.
  The advantage of helium over protons is significantly larger precision of the treatment due to lower
  scattering of alpha particles in the body, what is especially important for pediatric patients.
  A facility with therapeutic beams limited to helium, would be smaller and
  cheaper than carbon-based facility. Here we discuss this option.

\end{abstract}

\thispagestyle{empty}


\newpage

\section{Introduction}

\begin{figure}[!bht]
    \centering
    \includegraphics*[width=0.48\textwidth]{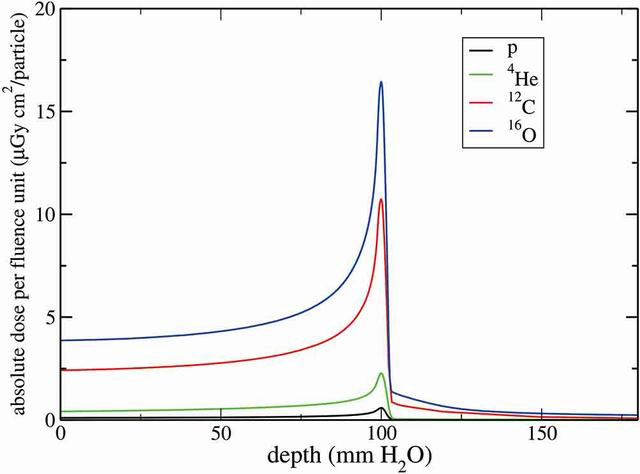} \hfill
    \includegraphics*[width=0.48\textwidth]{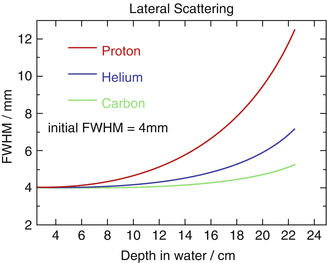}
    \caption{Comparison of properties of the ions used in cancer radiotherapy.
      Left plot shows absolute longitudinal dose profile. Lack of fragmentation tail is visible for protons and helium. Right plot shows the width of the lateral dose fall-off due to multiple scattering.
      Figures from \cite{Tomassino2016} (left) and \cite{Boldeman2010} (right plot).}
    \label{fig1}
\end{figure}

In comparison to protons, the helium ions provide reduced lateral scattering and enhanced biological damage (relative biological effectiveness, RBE) to tumors.
At the same time the particle fragmentation effects in distal healthy tissues are much smaller than
for heavier ions like carbon.
Multiple experiments with helium beams have been carried out in various facilities \cite{Knaus2016}.
The enhancement of RBE with respect to protons is moderate \cite{Mein2019}, however the precision of the
helium beam due to reduction of lateral scattering is significantly improved \cite{Tessonnier2017}.
This is an important factor for pediatric patients, where tumors are small and sensitive organs are
very close to the tumor.
The reduction of the fragmentation tail with respect to heavier ions is illustrated on the left plot of Figure \ref{fig1}. The comparison of lateral scattering, shown on the right plot of the Figure \ref{fig1},
means increased precision of the dose delivery.

The energy of the ion beams required by the therapy are determined by their range in the body.
The irradiation procedures require the range in the body to be from 3 cm (shallow tumors)
to 33 cm (deep-seated tumors) what corresponds to $\rm ^4He^{2+}$ beam energies between 60 MeV/u and 250 MeV/u.
The relation between energy $E_k$ and range $R$ (position of the Bragg peak) has approximately a
form $R=\alpha\cdot E_k^p$, where $\alpha$ is a constants and $p$ is in the range 1.78-1.82.
The range tables for protons and helium ions can be found in \cite{10.1093/jicru/os25.2.Report49}.

The maximum $\rm ^4He^{2+}$ beam energy of 250 MeV/u corresponds to
magnetic rigidity of $\rm 4.85~Tm$ what is only about 70\% of $\rm 6.62~Tm$ required for carbon ion beams.
The 70\% scaling factor gives an approximate idea how the size and cost of the helium-beam facility compares to carbon-beam facilities.


\section{Current trends in accelerators for radiotherapy}

Presently, the particle radiotherapy facilities are dominated by cyclotrons for protons and by synchrotrons for
heavier ions. The main reason is that proton cyclotrons are very compact and easy-to-operate machines.
For instance the IBA S2C2 superconducting synchrocyclotron has a diameter of about 2.5 meters and weights
around 50 tons ~\cite{S2C2_2013}.
The synchrotrons and linacs are larger than cyclotrons, with the most compact proton synchrotron
- new Hitachi design - occupying a square of $5\times 5.2~m^2$ surface, without the injector \cite{Umezawa2013}
(see Figure \ref{fig2}).
  
One of the most bulky element of a radiation therapy facility is shielding,
which protects the patients, the staff and the public from the radiation produced due to beam losses.
The most troublesome component of this radiation are high energy neutrons, which are difficult to shield
because they go through multiple elastic scattering events before they are stopped, therefore they can 'leak' through cracks and openings
in the radiation shield.
This radiation is mainly generated in extraction process and by Energy Selection Systems (ESS).
The compact medical cyclotrons cannot use charge exchange extraction, which is very efficient and minimizes beam losses,
because negative ions dissociate in strong magnetic fields.
Therefore, the most common is extraction in which orbits of the last turns are excited using resonance,
 creating separation of the last turn orbit and peeling off the beam by electrostatic deflector.
Synchrotrons also use resonant extraction, but applied to the whole beam in the machine.
In both cases significant percentage (10-30\%) of the beam is lost.
However, the main source of beam losses are degrades of the ESS system, which is used to tune the beam energy.
Those degraders also deteriorate the beam quality.
The ESS is needed only for cyclotrons, because synchrotrons can vary their extraction energy.
Therefore, the synchrotrons usually require thinner shielding, however, due to their larger size,
a much larger space must be shielded.
Therefore, the proton synchrotrons still compete on the market with cyclotrons.

A relatively new technology on the proton therapy market are the high-frequency linacs, which offer a fast pulse-to-pulse energy
variation with frequency of \hbox{200 Hz} \cite{Ungaro:2314160}.
This is an emerging technology and one of those linacs is
currently under construction (AVO-ADAM, UK).

\begin{table}[b!]
 \begin{center}
  \caption{Comparison of the main features of particle therapy accelerators.}
   \label{table1}    
    \begin{tabular}{l|l|l|l}
                  &  cyclotrons  & synchrotrons     & linacs \\
      \hline  
      beam energy       &  fixed       &  variable        & variable \\
      beam structure    &  continuous   & spills (0.2-10 s) & pulses of ~$\rm 10~\mu s$ at 200 Hz \\
       \begin{tabular}{@{}c@{}}size (p,He,C)[m] \\ (diameter/length)\end{tabular} 
      &  2.5, 5.2, 6.9    & 5.5, 15, 20        & 24, 40, 54 \\
      technology state  & established  & established      & emerging \\
    \end{tabular}
 \end{center}
\end{table}

A summary of the technologies is presented in Table \ref{table1}. Note that comparison of sizes is not straightforward; the linacs are
much longer than synchrotrons and cyclotrons, but their width is much smaller.
In the Table the length of not folded linac is given.
The final footprint of facilities based on linacs is smaller than normal-conducting synchrotrons.

For ions heavier than protons, the advantage of synchrotrons over cyclotrons become more emphasized.
The larger maximum rigidities of the beams require much larger cyclotron magnets, therefore currently
there is only one project of carbon-cyclotron therapy machine \cite{Jongen:2008zz},
while the other facilities are based on synchrotrons.

A helium linac was proposed in \cite{Benedetti2018} but currently, to author's knowledge,
there are no projects focused on delivering helium-only therapeutic beams.
Because of medical advantages of helium over protons and significant reduction of cost and size of the facility
in comparison with carbon ions, a helium-beam cancer treatment facility is investigated.
All three types of accelerators can be considered, however,
because the linac technology is still emerging and the helium cyclotron is a massive device,
here an example of synchrotron-based facility is drafted.

\section{Example of helium-only facility}

The  proposed system is based on classical concept of injector linac and synchrotron,
but makes use of exciting recent developments.
The crucial development is the injector, made of 750 MHz RFQ developed at CERN, which is the most compact RFQ developed to date.
This RFQ could be used as an injector alone, without additional accelerating structures.
This was already suggested in the 1991 \cite{Becker1991}, however, at that time the RFQ output energies were too low for efficient injection into synchrotron.
The modern RFQ can provide higher beam energy and the direct injection to the synchrotron, without additional accelerating structures, should be reconsidered.
The use of RFQ direct injection alone makes a significant difference in the complexity and price of an ion therapy facility.

\subsection{Sources and injector}

The beam is produced by two Electron-Beam Ion Sources (EBIS).
Those sources produce beams with much smaller emittance than
the Electron-Cyclotron Resonance Ion Sources (ECRIS) used in ion-therapy facilities.
Due to lower pressure EBIS sources generate less impurities what particularly important as from the beginning the accelerators are tuned to charge-to-mass ratio of 1:2, what could lead to many other ions being accelerated, for instance $\rm H_2^+, ^{12}C^{6+} or~ ^{16}O^{8+}$.
The main disadvantage of EBIS source is a low beam current. 
For instance a compact Dresden-EBIS-SC source can produce only $\rm 2\cdot 10^9$ $\rm ^{4}He^{2+}$ ions per pulse.
The pulse length can be regulated in a broad range  $\rm 0.1~to~ 100~\mu s$.
The total number of particles, which are needed in the synchrotron to treat $\rm 2~dm^3$ tumor,
is about $\rm 8\cdot 10^{10}$, what corresponds to 40 pulses, however the transmission through the injector is only about 50\%.
Multi-turn injection over such large number of turns is challenging.
Nevertheless these parameters are close to the requested ones and
ongoing developments give hope that EBIS sources can replace ECRIS in medical facilities in the next few years \cite{Shornikov:2255889}.
Two of such sources should be installed to allow for fast switching in case one of them fails.

The typical injector linac is a complex system with RFQ and IH structures and accounts for about 30\%
of the cost of the facility.
The 750 MHz RFQs features small cross section of the cavity
and innovate beam dynamics which allow for a very efficient acceleration \cite{Lombardi:IPAC2015-WEYB2}.
One of these RFQs, designed for protons, has been installed on LIGHT proton therapy linac.
The design based on similar assumptions has been prepared for $\rm C^{6+}$ ions \cite{Bencini:2669960} and
the same design could be used for $\rm ^4He^{2+}$ beam.

This design foresees the output energy of \hbox{2.5 MeV/u} or \hbox{5 MeV/u}.
Especially in the latter case, the RFQ could be the only accelerating element of the injector linac.
The length of the RFQ is \hbox{2.7 m} for output energy of 2.5 MeV/u or \hbox{5.8 m} for \hbox{5 MeV/u}.
The injection energy to the synchrotron at 5 MeV/u is close to the optimal
defined by efficiency of multi-turn injection.
Probably slightly lower injection energy should be studied.
Here the output energy of \hbox{4 MeV/u} and the length of \hbox{4 m} are assumed. No additional IH structure is needed and therefore,
the cost of the system should be reduced by about 50\% with respect to contemporary carbon injector linacs.

The transfer line between the injector and synchrotron includes a debuncher cavity in order to decrease the momentum spread of the beam and to facilitate the RF-capture in the synchrotron.

\begin{table}[h!]
 \begin{center}
  \caption{Parameters of the injector linac}
   \label{table2}    
   \begin{tabular}{l|l}
     \hline
      Source                     & Twin-EBIS \\
      Source current             &  $\rm 2\cdot 10^9$ particles per pulse \\  
      Pulse duration             &  $\rm 100~ns-100~\mu s$ (verify)\\
      RFQ frequency              &  750 MHz \\
      RFQ Transmission           &  50\%\\
      Overall length             &  about 7 m   \\
      Final energy               &  4 MeV/u \\
    \end{tabular}
 \end{center}
\end{table}

\subsection{Synchrotron}

Numerous designs of carbon and proton synchrotrons have been developed.
Scaling up or down one of these machines should provide a good design for helium machine.

\begin{figure}[!tbh]
    \centering
    \includegraphics*[width=0.65\textwidth]{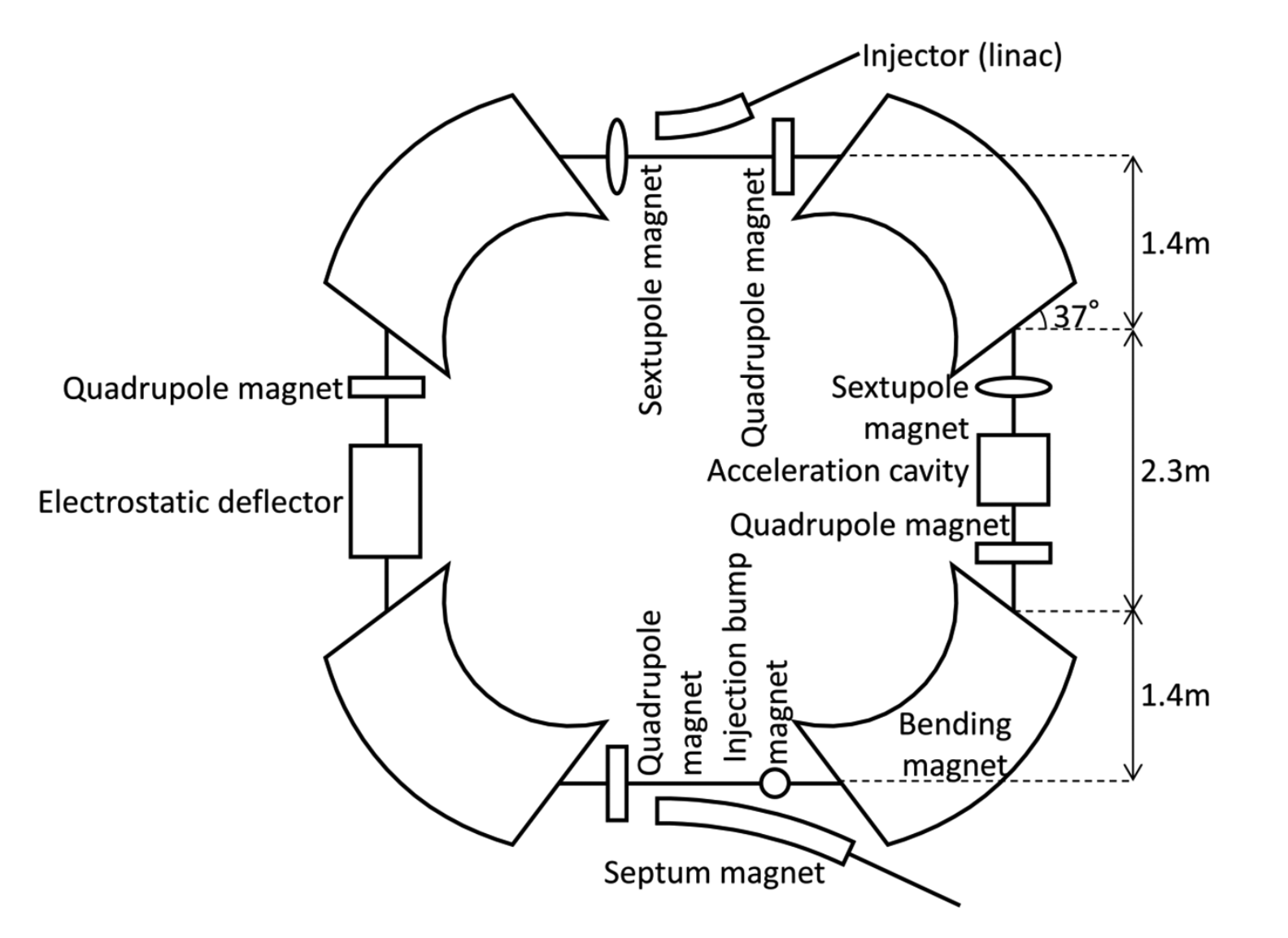}
    \caption{Hitachi proton synchrotron. Figure from \cite{doi:10.1002/ecj.11972}}
    \label{fig2}
\end{figure}

The current commercial proton synchrotrons often use four 90\textdegree~ main dipoles
(for example Hitachi design, see Figure \ref{fig2}, ProTom splits them into 8 keeping the square shape)
to reduce its footprint and fit into typical rectangular rooms.
Very compact designs usually relay on edge focusing to reduce number of quadrupoles and suffer from lack of dispersion-free
sections, but the circumference of the machine is as small as 18 m \cite{doi:10.1002/ecj.11972} or ever 16 m \cite{Vostrikov:2018uko}.

Heidelberg Ion Therapy center (HIT) synchrotron has 6 main dipoles, what is the smallest number among carbon machines.
Circumference of this machine is about 65 meters but dipoles are very heavy and there are no have dispersion-free sections.


The progress in synchrotron design in the last 20 years comes from synchrotron light sources.
Those machines are optimized for minimum beam size and therefore their lattices provide the smallest $\rm \beta$ functions.
Those lattices are multi-bend achromats.
The small beam size is also useful for hadron therapy machine, as the small vertical
aperture leads to cheaper and lighter dipoles, which are one of the cost drivers of the facility.
Therefore, a carbon-therapy machine based on double-bend achromat lattice has been proposed recently \cite{zhang2020lattice}.
This preliminary design features also a reduced-size of machine (55 m of circumference) and dispersion-free regions, which are ideal
locations for RF cavities and injection/extraction devices.

In this exercise an unconventional lattice made of double-bend achromatic (DBA) and triple-bend achromatic sections (TBA) is investigated. Five main dipoles is probably the smallest number of bends which
can be reasonably used to construct a synchrotron at the required rigidities. 

A warm synchrotron can typically operate with bending fields up to \hbox{1.5 T},
therefore, for the maximum required rigidity of helium beam $\rm B\rho = 4.85~Tm$ the bending radius of the dipoles is only \hbox{3.23 m}.
The length of these magnets is around \hbox{4.1 m}.
These are large magnets, however they should be compared with 60\textdegree~ dipoles installed in HIT synchrotron, with length of \hbox{4.6 m} and mass of 23.5 tons.

\begin{figure}[!bht]
    \centering
    \includegraphics*[width=0.48\textwidth]{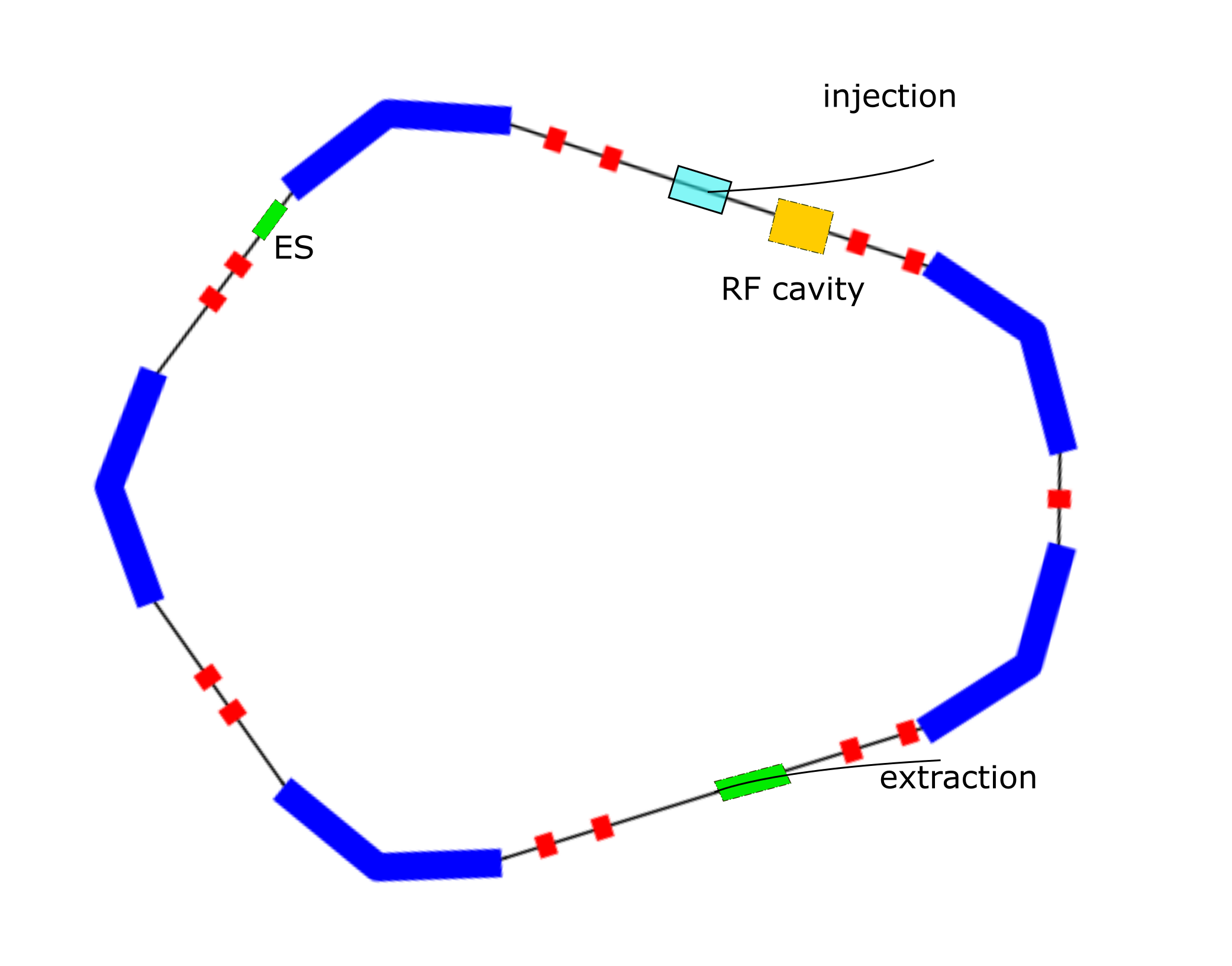}
    \includegraphics*[width=0.48\textwidth]{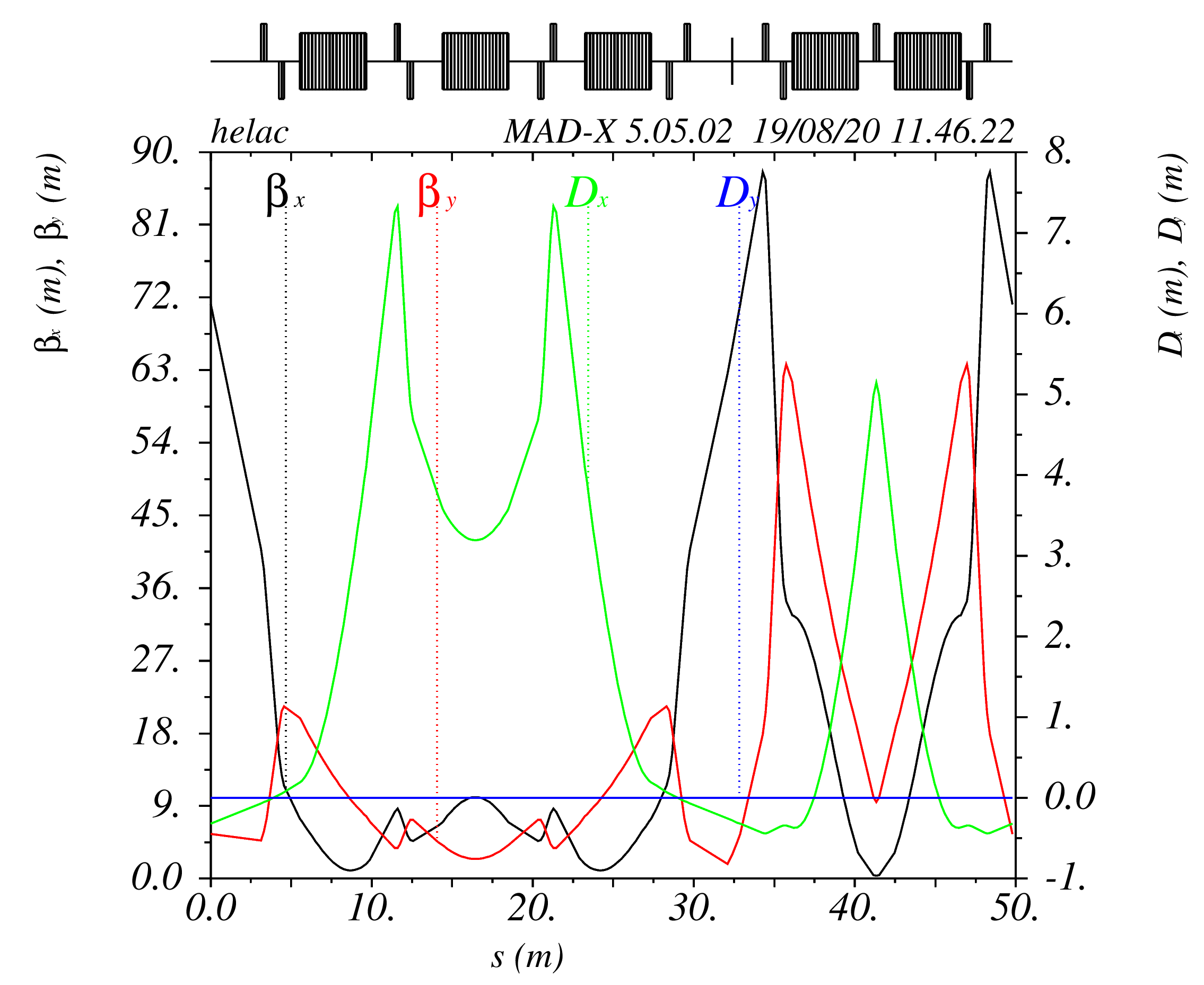}
    \caption{Left plot shows the layout of helium synchrotron. Right one shows the synchrotron lattice.}
    \label{fig3}
\end{figure}

The synchrotron was designed using MAD-X \cite{MADX}.
The layout of the machine is shown in Figure \ref{fig3}. 
It has a form of a pentagon with two sides elongated.
Those are dispersion-free regions, each about 5 meters long, which can fit comfortably 
injection and extraction equipment and RF-cavity.
The circumference of the whole machine is 50 meters.

\begin{table}[h!]
 \begin{center}
  \caption{Parameters of the proposed synchrotron for helium particle therapy beams.}
   \label{table3}    
   \begin{tabular}{l|l}
     \hline
      Tunes $Q_H, Q_V$                  &  1.68, 1.59 \\
      Maximum $\beta_H, \beta_V$        &  88, 64 m \\
      Natural chromaticity               & -9.65, -8.15 \\
      Transition $\gamma$              &  2.755    \\
      Dipole strength                  & 1.5 T \\
      Dipole length                    & 4.1 m \\
      Quadrupole length                & 0.34 m \\
      Circumference                    & 49.79 m \\  
      Injection energy                 & 4 MeV/u \\
      Extraction energy                & 70-250 MeV/u \\
      Maximum rigidity                 &  4.85 Tm \\
      Revolution period at injection &  $\rm 1.8~\mu s$\\
      Revolution period at extraction & $\rm 0.27~\mu s$ \\
    \end{tabular}
 \end{center}
\end{table}

In this preliminary design there are 13 quadrupoles in 7 families. Sextupoles must be installed
in dispersive regions to control the chromaticity.
The advantage of low $\beta$ is seen only in the TBA part of the lattice, while in the dispersion-free region $\beta$s reach high values what helps to increase the separation of particles being on their last turn before extraction.

The RF system consists of 0.5-3.7 MHz Finemet® loaded wideband cavity powered by solid state amplifiers, similar to the systems installed in carbon ion facilities.
They are state of the art cavities for low energy accelerators \cite{Ohomori2005}
and allow operation in multi-harmonics mode.

The slow extraction is done by RF-KO method, which does not require beam debunching and therefore simplifies Multi-Energy Extraction, what is currently considered as one of the most important improvements in operation of carbon ion facilities.
The electrostatic septum (ES) is about 60-cm long and is installed in a 2-meter drift space between main dipole and quadrupole TBA section.
The phase advance between ES and magnetic septum is 267\textdegree.
It is strongly dependent on the optics.
In order to fulfill Hard't condition the additional sextupoles are installed in dispersion-free sections.

The lattice is preliminary and there is space for improvements.
It could profit stronger from weak focusing or use combined function magnets and the dipoles could be split to more manageable size.
The final design could be 10-15\% more compact. The main parameters of the proposed synchrotron
are summarized in Table \ref{table3}.

\subsection{Transfer lines and gantry}

Transfer lines connect the extraction from the synchrotron with the patient.
The beam can be delivered using a fixed beam line but many medical treatment plans
foresee use of a gantry.
Proton gantries are common, commercially available products.
Typical gantry has about 3.5 meters of diameter and weights around 100 tons.
Carbon gantries are much larger and currently only two such objects exist:
one based on normal conducting magnets at HIT (Germany) \cite{Galonska2011}
and one based on superconducting magnets at NIRS (Japan) \cite{Iwata:IPAC2018-TUZGBF1}.
New projects of superconducting carbon gantries are ongoing.
Helium gantry could be a scaled down version of one of these new gantries.

\begin{figure}[!bht]
    \centering
    \includegraphics*[width=0.75\textwidth]{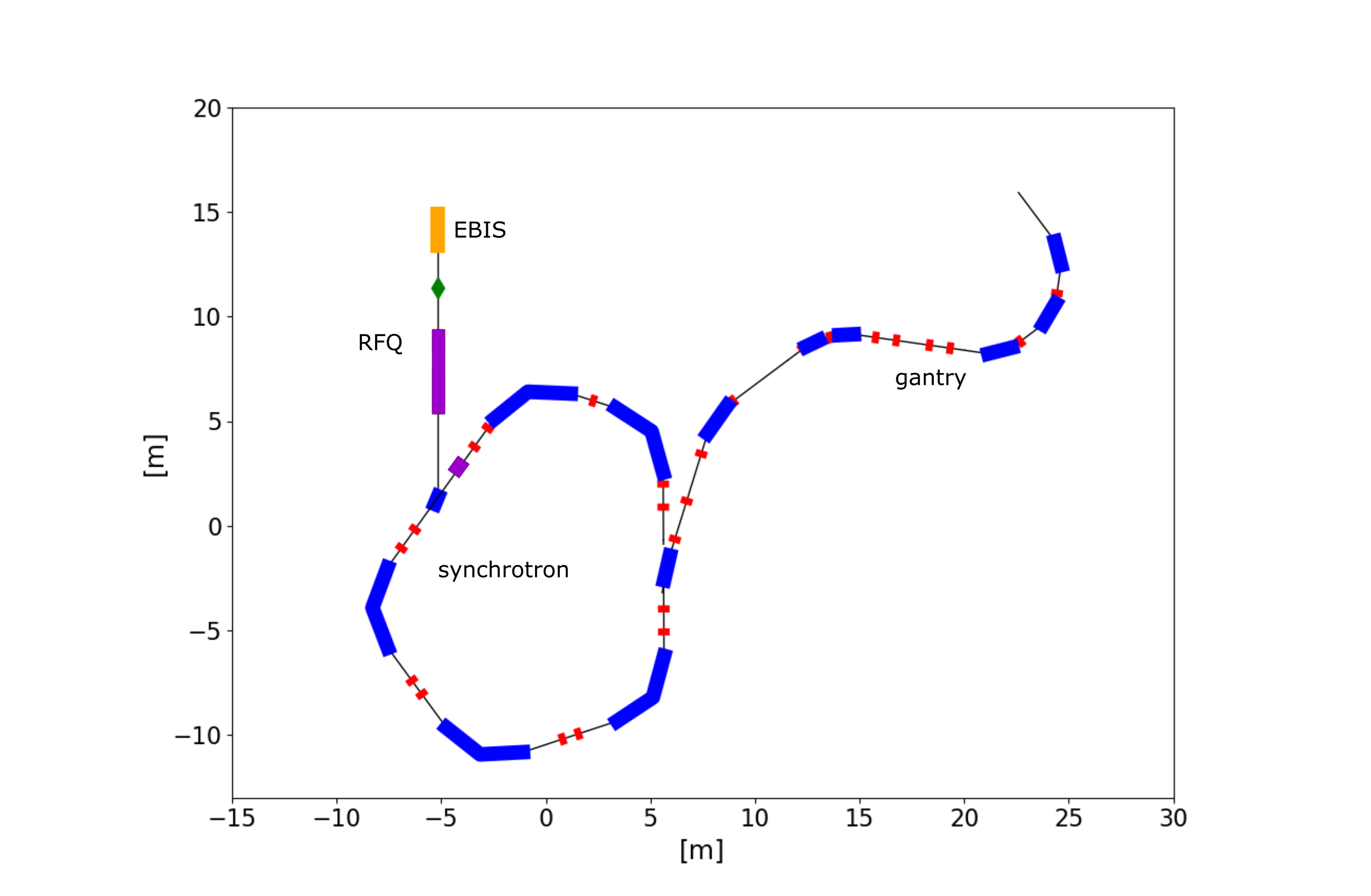}
    \caption{The single-room layout of helium irradiation facility.}
    \label{fig4}
\end{figure}

Figure \ref{fig4} shows the whole facility in a single room configuration with a gantry.


\section{Conclusions}

The goal of this article is to build a case for helium ion beam therapy facility.
Helium beams offer a significant increase in precision of the irradiation process over protons.
With the final cost of about two to three times of a synchrotron based proton therapy systems,
it is an interesting alternative, which could be commercialized easier than carbon ion therapy systems.
Helium-beam therapy can complement the carbon therapy but
will not replace it because radioresistant tumors require high RBE provided only by carbon ions.

The particular solution discussed here is a conceptual design of the synchrotron
based on double and triple-bend achromats connected by spacious dispersion-free regions.
The big advantage is an injector made of RFQ only, what greatly simplifies the system.
Development of such a machine should be accompanied by development of EBIS ion sources,
which produce beams with much smaller emittance than currently used ECR sources. 

\bibliography{helac,long}{}

\bibliographystyle{ieeetr}

\end{document}